# The Imprint of Proper Motion of Nonlinear Structures on the Cosmic Microwave Background

Robin Tuluie[1] and Pablo Laguna[2]
*Department of Astronomy and Astrophysics*
*and*
*Center for Gravitational Physics and Geometry*
*The Pennsylvania State University*
*University Park, PA 16802*

## ABSTRACT

We investigate the imprint of nonlinear matter condensations on the Cosmic Microwave Background (CMB) in an $\Omega = 1$, Cold Dark Matter (CDM) model universe. Temperature anisotropies are obtained by numerically evolving matter inhomogeneities and CMB photons from the beginning of decoupling until the present epoch. The underlying density field produced by the inhomogeneities is followed from the linear, through the weakly clustered, into the fully nonlinear regime. We concentrate on CMB temperature distortions arising from variations in the gravitational potentials of nonlinear structures. We find two sources of temperature fluctuations produced by time-varying potentials: (1) anisotropies due to intrinsic changes in the gravitational potentials of the inhomogeneities and (2) anisotropies generated by the peculiar, bulk motion of the structures across the microwave sky. Both effects generate CMB anisotropies in the range of $10^{-7} \lesssim \Delta T/T \lesssim 10^{-6}$ on scales of $\sim 1°$. For isolated structures, anisotropies due to proper motion exhibit a dipole-like signature in the CMB sky that in principle could yield information on the transverse velocity of the structures.

*Subject headings:* cosmic microwave background – cosmology

---

[1] e-mail: rtuluie@astro.psu.edu
[2] e-mail: pablo@astro.psu.edu

Recently, there has been a renewed interest on studying anisotropies of the CMB produced by the late time evolution of density inhomogeneities (Martínez-González and Sanz 1990; Martínez-González et al. 1990; Martínez-González et al. 1992; Panek 1992; Fang and Wu 1993; Meszáros 1994; Martínez-Gonzá.lez et al. 1994; Hu and Sugiyama 1994;). However, most of these and other studies in the past (Rees and Sciama 1968; Dyer 1976; Kaiser 1982; Thompson and Vishniac 1987; Dyer and Ip 1988; Nottale 1984) have only considered the effects on the CMB by a single, spherically symmetric structure using Swiss cheese models, thin-shell approximations or Tolman-Bondi solutions. The imprint on the CMB of multiple, non-trivially distributed matter structures, as well as the unambiguous linkage between the linear, weakly clustered and fully nonlinear epochs, has not been yet taken into consideration; in particular, the peculiar motion of matter structures has been ignored so far.

In this *letter*, we address the signatures in the microwave sky from time-varying gravitational potentials associated with nonlinear density inhomogeneities in the universe, first studied by Rees and Sciama in 1968. This effect, hereafter referred as the Rees-Sciama effect, is a secondary source of temperature anisotropy; however, it could yield important information on the underlying density field that generates this effect. Our goal is to study the contribution of the Rees-Sciama effect to the total CMB anisotropy, in particular, finding the correlation between the Rees-Sciama effect with the underlying density field and proper motion of structures that photons encounter along their paths. For this reason, we have ignored Doppler shift anisotropies from Thomson scattering. Doppler shift anisotropies from late Thomson scattering are important if a substantial reionization of the IGM occurred (Tuluie et al. 1995).

Our study is based on numerical simulations of unbiased, $\Omega = 1$, $h = 0.75$, CDM models of the universe. For the parameters of our models, voids are typically of size $\sim 30h^{-1}$ Mpc, and structures producing the stronger signatures $\sim 10h^{-1}$ Mpc. Our study follows the approach developed by Anninos et al. 1991 that relates density inhomogeneities to CMB fluctuations by evolving CMB photons through the forming matter inhomogeneities. It is important to emphasize that no assumptions are made about the nonlinear nature of the density perturbations or peculiar bulk motions. We only require sub-horizon scales and non-relativistic proper velocities, so linear theory of cosmological perturbations holds (Mukhanov et al. 1992). Furthermore, since we propagate CMB photons through matter that evolves from the beginning of decoupling up to present time, our results naturally take into consideration not only the gravitational lensing of the CMB caused by the late, nonlinear density perturbations, but also the cumulative effects on the CMB photons as they propagate through multiple and different regions of voids and clusters.

In linear theory of cosmological perturbations (Mukhanov et al. 1992), the evolution of the density field is governed by a gravitational field $\phi$ which represents small perturbations against the homogeneous background ($\phi \ll 1$). The temperature anisotropies of the CMB associated with the gravitational potential $\phi$ are given by (Sachs and Wolfe 1967)

$$\frac{\Delta T}{T} = \phi_r - \phi_e + 2 \int_r^e \frac{\partial \phi}{\partial t} dt \qquad (1)$$

where the integral is taken along photon trajectories. The first and second terms in equation (1) represent the Sachs-Wolfe and Rees-Sciama effect, respectively. The numerical factors multiplying each term in equation (1) differ among various authors depending on the choice of gauge (Padmanabhan 1993). Our results were obtained using the longitudinal gauge only. In an $\Omega = 1$ matter-dominated universe, like the one here considered, the gravitational potential $\phi$ is static as long as the density perturbations are linear. Consequently, in the linear regime the Rees-Sciama effect is absent. In our model, the nonlinear regime is reached at $z \sim 3$; nonetheless, the underlying gravitational field is still well described by linear theory ($\phi \ll 1$). On sub-horizon scales, this implies that the system can still be treated with Newtonian gravity in comoving coordinates (Mukhanov et al. 1992). During the nonlinear epoch, the gravitational field can become significantly non-static due to the evolution of the density field. CMB photons, which mirror the gravitational field $\phi$, can thus reflect not only the signature of the gravitational potentials at the surface of last scattering, namely the Sachs-Wolfe effect (Sachs and Wolfe 1967), but also the imprint from nonlinear density perturbations at low redshifts.

Our technique for calculating CMB temperature anisotropies proceeds in two stages (see Anninos et al. 1991 for a detailed description of the numerical scheme and tests). First, starting at a redshift of $z = 1900$, just prior to recombination, a $64^3$ particle-mesh model is used to simulate the CDM from an initial Harrison-Zel'dovic spectrum modified by the appropriate damping term. Since only sub-horizon scales are considered, the collisionless matter is evolved using Newtonian gravity in an expanding grid. The simulation evolves from the linear, through the weakly clustered, into the nonlinear regime and terminates at z=0. The initial amplitude of the perturbations is the only free parameter in the model and was chosen so that the r.m.s. overdensity in $8h^{-1}$Mpc scales, $\sigma_8$, reaches unity at z=0. We have verified that this normalization yields both reasonable large scale structure and peculiar velocity fields for our simulations. The results presented here are from CDM models in a computational grid $124h^{-1}$ Mpc on a side, so the minimum resolution is $\sim 2h^{-1}$ Mpc. In the second stage of our method, a bundle of photons is propagated back in time through the evolving dark matter structures, starting at at z=0 and finishing at z=1900. We propagate bundles with up to $10^6$ individual CMB photons, updating their individual temperature at every timestep by computing explicitly the Rees-Sciama effect from the second term in equation (1).

The correlation between the Rees-Sciama effect and density inhomogeneities are performed via the column density contrast of dark matter along the photon trajectories. This column density contrast is defined as

$$\sigma \equiv \int_r^e \frac{\delta\rho}{\rho} dt, \qquad (2)$$

where the integral is along the photon trajectories. Last, we construct maps of $\Delta T/T$(Rees-Sciama) and the column density contrast $\sigma$.

Figure 1 shows the r.m.s. value of the Rees-Sciama effect as a function of angular scale of the photon bundle. The lower limit of the dynamical range of the simulation, which is $\sim 0.1°$, is dictated by the finite resolution of our computational model. The upper limit of the dynamical range, $\sim 8°$, is set to avoid boundary effects from our periodic computational domain. The error bars indicate the cosmic variance from 42 simulations, that is, the variance in the Rees-Sciama effect for different directions at a fixed angular scale. Typically, these variances are below the 15% level for photon bundles with angular scales $> 2°$. For a particular photon bundle direction on angular scale $2°$, figure 2 shows the accumulated r.m.s. values of the Rees-Sciama effect (*top*) and column density contrast $\sigma$ (*bottom*) as a function of redshift. Other directions were considered, and a similar outcome was obtained. Three features are evident from figure 2. First, as expected, the Rees-Sciama is produced almost entirely during the nonlinear stage of the density fields ($z \lesssim 3$). Second, increases in the accumulation of the Rees-Sciama effect are correlated with increases in the column density contrast. These jumps are due to high density structures within the photon beam. Third, because comoving voids make the CMB photon appear warmer while high density structures make the photon appear colder (Dyer and Ip 1988), the accumulation of the Rees-Sciama effect is not exactly monotonic; clusters, however, dominate the Rees-Sciama signal due their high density contrast ($\delta\rho/\rho \gtrsim 10$) compared to our $\sim 30$ Mpc voids ($0 \gtrsim \delta\rho/\rho \gtrsim -1$), which are still in the quasi-linear regime.

The comparison between the average density contrast $\delta\rho/\rho$ and the r.m.s. change of the gravitational potential $\frac{\partial\phi}{\partial t}$ within the photon bundle for $z < 0.1$ is depicted in figure 3. Here the voids, clusters and filaments that the photon beam traverses are directly mirrored in the Rees-Sciama signal through $\frac{\partial\phi}{\partial t}$. We see clearly from this figure that indeed overdense structures provide the major contribution to the Rees-Sciama effect.

Figures 4 through 7 are maps from the propagation of $10^5$ photons as seen by a comoving observer at $z = 0$. Figures 4 and 5 show respectively the Rees-Sciama effect and column density contrast $\sigma$ up to $z = 1900$ of a $2° \times 2°$ photon beam. Figure 5 represents a map of the total line-of-sight density fluctuations, i.e., the galaxy clusters and voids "seen" by the CMB photons in figure 4. The poor map correlation between figures 4 and 5 is due to the smearing by the void-cluster-void-cluster cumulative Rees-Sciama fluctuations. Figure 6 and 7 show again the Rees-Sciama effect and column density contrast $\sigma$, respectively. In this case however, the maps are for several isolated structures $\sim 10h^{-1}$ Mpc across using a $16° \times 16°$ photon beam. The slab is located at $0.065 < z < 0.095$, about $330h^{-1}$ Mpc away from the observer and nearly $110h^{-1}$ Mpc thick. This is the region that yields the maximum density contrast in figure 3. We have also examined the region around $z = 0.03$ that produces the maximum $\frac{\partial\phi}{\partial t}$, and maps with similar features were obtained. Since the integration to generate figures 6 and 7 is done through a volume thin enough, smearing effects are avoided in this case. There is a clear correlation between the Rees-Sciama effect (Fig. 6) and the column density contrast (Fig. 7), showing that indeed the Rees-Sciama effect is caused by the matter structures. In order to quantify the correlation between two given maps, we define the correlation strength of maps A and B by

$$\xi^2 = \frac{1}{N} \sum_{pixels} \frac{(a_{ij} \times b_{ij})}{(a_{ij}^2 + b_{ij}^2)/2}, \qquad (3)$$

where $a_{ij}$ and $b_{ij}$ are the corresponding pixel values of maps A and B normalized to unity, and $N$ the total number of pixels. We found a correlation strength $\xi = .71$ between figures 4 and 5. In contrast, figures 6 and 7 have a stronger correlation of $\xi = .93$.

A more careful examination of the maps in figures 6 and 7 shows that the location of the high density structures in figure 7 are not centered at the cold (*dark*) spots in the temperature map (Fig. 6) as one would expect. Furthermore, most of the dark spots in the temperature map that are located near high density structures have associated with them a hot (*white*) spot, with the structure centered between adjacent hot and cold spots. This configuration resembles a hot-cold temperature dipole. These dipole-like patterns in temperature fluctuations are due to the transverse peculiar bulk motion of the cluster and its associated gravitational potential. Photons crossing the path of the cluster ahead of the approaching cluster will be redshifted because of the increasing depth of the potential well during the photon crossing time, while photons crossing at the trailing end of the structure are blueshifted. Hence, even an intrinsically static potential can cause a significant temperature anisotropy as long as the transverse proper motion and mass of the cluster are large enough. The separation between dipole peaks is indicative of the characteristic length-scale of the density perturbation; furthermore, the temperature difference between the dipole peaks is a measure of the transverse velocity of the cluster. A detailed analysis of these dipole patterns in the CMB sky due to proper motion of the inhomogeneities will be presented elsewhere (Laguna and Tuluie 1995).

One can get an estimate of the total effect from proper motions and intrinsic potential variations from the following scaling argument. The contribution to the temperature anisotropy due to an isolated structure is

$$\frac{\Delta T}{T} \sim \frac{\partial \phi}{\partial t} \delta t \sim \left( \frac{v}{d} \phi + \frac{\phi}{t_c} \right) \delta t \sim v\phi + \phi^{3/2}, \qquad (4)$$

where $\delta t$ is the photon crossing time, $v$ is the peculiar velocity of the structure, $d$ its physical size ($d \sim \delta t$) and $t_c \sim d/\phi^{1/2}$ its characteristic evolution time (Panek 1992). The first term in equation (4) arises from proper motions and the second from intrinsic changes in the potential. On the other hand, from linear perturbation theory we have that $v \sim \phi(1+z)^{-3/2}(d\,H_o)^{-1}$ and $\delta\rho/\rho \sim \phi(1+z)^{-3}(d\,H_o)^{-2}$, so equation (4) can be rewritten as

$$\frac{\Delta T}{T} \sim 10^{-7} \left[ \left( \frac{\delta\rho}{\rho} \right)^2 + \left( \frac{\delta\rho}{\rho} \right)^{3/2} \right] \left( \frac{d\,(1+z)^{3/2}}{14h^{-1}Mpc} \right)^3. \qquad (5)$$

Within the limitations of this simple argument, we see from equation (5) that the effect from proper motions is as important as that arising from intrinsic changes in the gravitational potential of the structures; moreover, if the cluster or supercluster under consideration is highly nonlinear, the effect from proper motions could have the dominant role in generating these secondary perturbations to the primeval CMB.

In summary, our study considers the gravitational influence of the evolving density field on the CMB directly from the dark matter dynamics and does not rely on any assumed symmetry of the structures. The underlying matter structures used exhibit the sheet-filament-knot structure of voids, clusters and superclusters, providing then a more realistic representation of the shape of density inhomogeneities in the universe. Furthermore, our numerical evolution avoids artificial splitting between linear, weakly clustered and nonlinear epochs. Our simulations indicate that the Rees-Sciama effect is small, $10^{-7} \lesssim \Delta T/T \lesssim 10^{-6}$, for void scales $\sim 30 - 60h^{-1}$ Mpc and clusters and superclusters $\lesssim 15h^{-1}$ Mpc. Our results disagree with those from authors who use larger estimates for the scale and density contrast of large scale matter structures (Meśzarós 1994; Rees and Sciama 1968). For the scales of the structures we considered ($\sim 10h^{-1}$ Mpc), we found that the Rees-Sciama effect, in addition to being produced by the intrinsic evolution of the density field (e.g. changes in the depth of the potential wells), is created by the change of the potential as the cluster moves across the microwave sky. We have found that this effect can be as important as that from the intrinsic evolution of the gravitational potentials in generating second-order temperature anisotropies. Finally, our simulations show that $\Delta T/T$ anisotropies induced by peculiar velocities leave a dipole-like imprint.

We thank P. Anninos, J. Carlton, R. Matzner, T. Padmanabhan and D. Spergel for numerous discussions and helpful suggestions. Work supported in part by NSF Young Investigator award PHY-9357219, NSF grant PHY-9309834 and NASA (at Los Alamos National Laboratory).

---

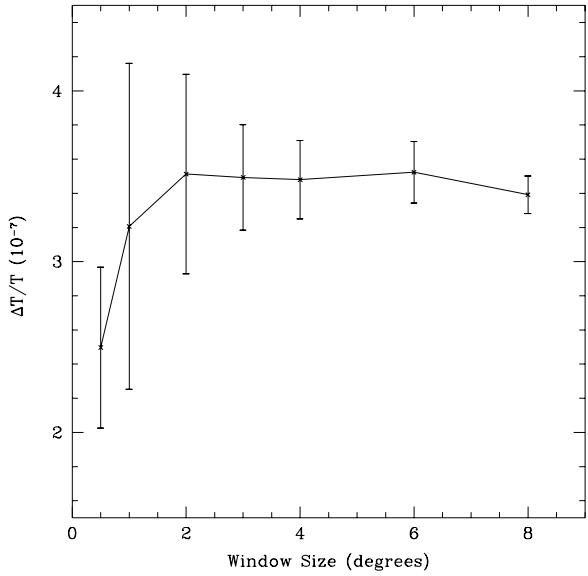

Fig. 1.— CMB $\Delta T/T_{rms}$ from the Rees-Sciama effect as a function of angular scale. The error bars are the cosmic variance from 42 simulations using $10^5$ photon beams. The minimum and maximum angular scales are set by the resolution of the simulations $\sim 2h^{-1}$ Mpc and the periodicity of the computational domain, respectively.

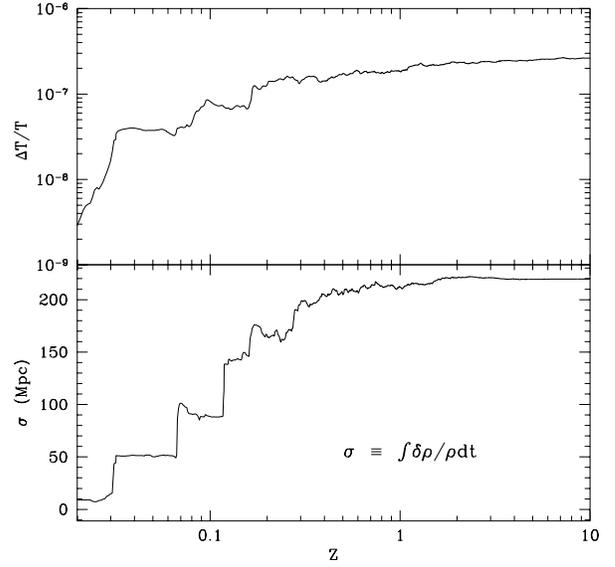

Fig. 2.— Accumulated r.m.s. anisotropy of the Rees-Sciama effect (*top*) and average column density contrast $\sigma$ (*bottom*) as a function of redshift. The results were obtained from a $10^5$ photon bundle with angular scale of $\sim 2°$. As expected, the Rees-Sciama effect increases significantly during the nonlinear stage of the matter evolution $z \leq 3$. The contribution from overdense regions is more dominant, so complete cancellation between dense regions or voids is not observed.

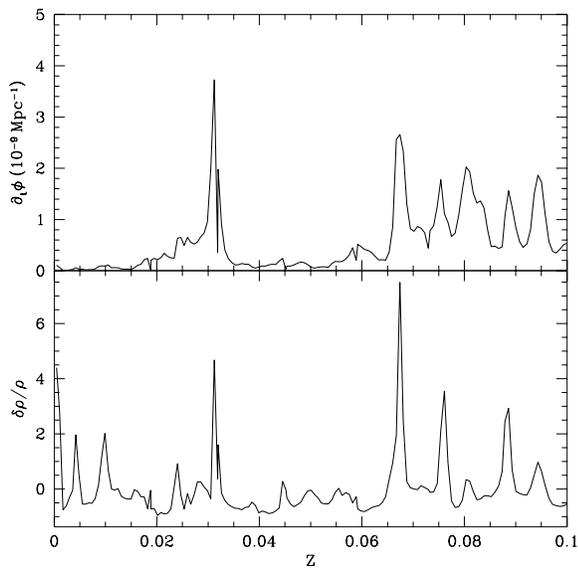

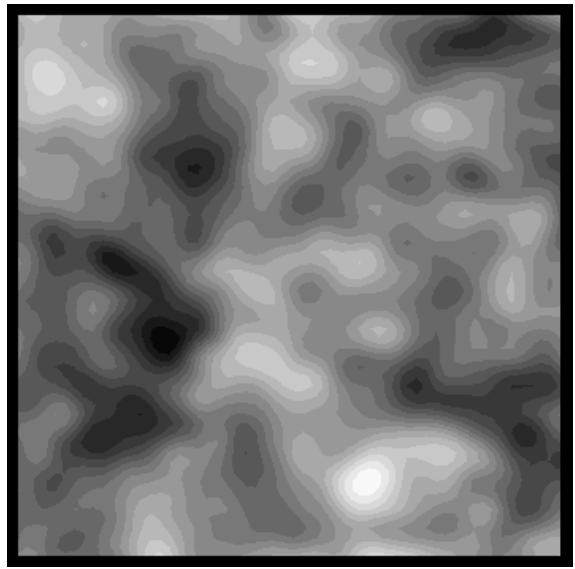

Fig. 3.— Comparison of $\frac{\partial \phi}{\partial t}$ r.m.s. (*top*) with the average density contrast $\delta\rho/\rho$ (*bottom*). There is a direct correlation between the voids and clusters that the photon beam traverses with the rate of change of the potentials ($\frac{\partial \phi}{\partial t}$). Overdense structures provide the major contribution to the Rees-Sciama effect. The results were obtained from a $10^5$ photon bundle with angular scale of $\sim 2°$.

Fig. 4.— Temperature map of $10^5$ photons in a randomly chosen direction of the simulated microwave sky for a $2° \times 2°$ window showing the Rees-Sciama effect only. For this and subsequent maps, maximum and minimum values are set black (cold) and white (hot), respectively. Here, $\Delta T/T_{rms} = 2.9 \times 10^{-7}$ and $-8.9 \times 10^{-7} \leq \Delta T/T \leq 9.5 \times 10^{-7}$.

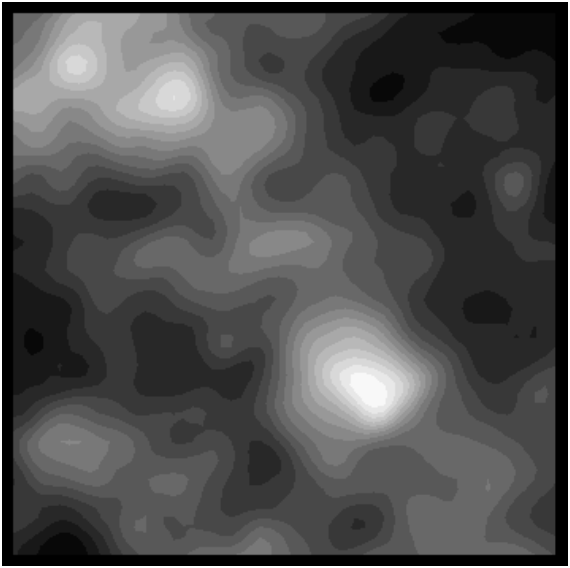
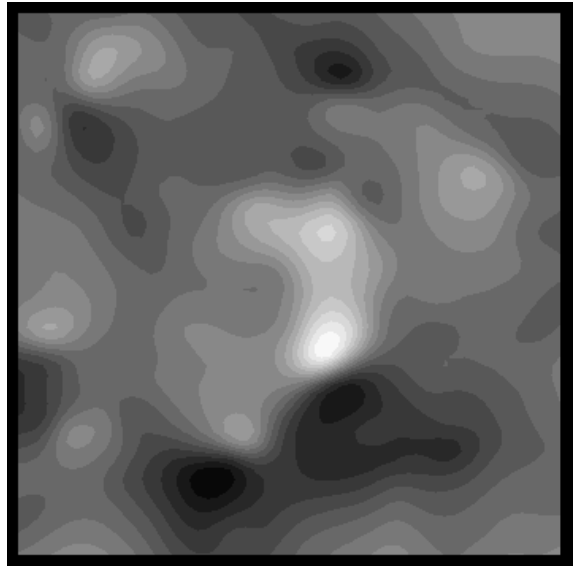

Fig. 5.— Map of the column density contrast, $\sigma$, for figure 4. This map provides the line-of-sight column density contrasts seen by the photons. Here, $\sigma_{avg} = -9.5$ Mpc and $-398$ Mpc $\leq \sigma \leq 822$Mpc.

Fig. 6.— Map of the Rees-Sciama effect for an isolated region $\sim 110h^{-1}$ Mpc thick located at $z \sim 0.08$, containing several typical clusters and voids. The angular size of the window in this case is $16° \times 16°$. Here, $\Delta T/T_{rms} = 7.6 \times 10^{-8}$ and $-2.6 \times 10^{-7} \leq \Delta T/T \leq 3.7 \times 10^{-7}$. Most of the anisotropies are due to the dominant clusters seen in figure 7. The temperature fluctuations have dipolar features produced by the proper motions of the clusters across the sky.

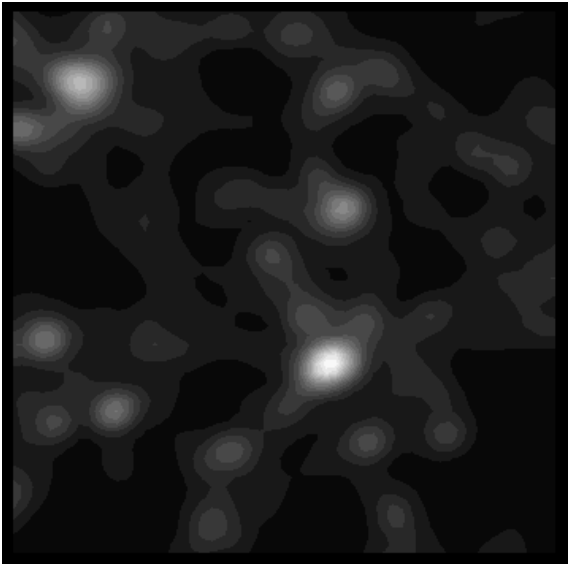

Fig. 7.— Column density contrast, $\sigma$, in dimensionless units for the localized region in figure 6. High density regions lie at the center of the dipolar patterns in figure 6. Here, $\sigma_{avg} = 16$ Mpc and $-70$ Mpc $\leq \sigma \leq 832$ Mpc.